# Using Paxos to Build a Scalable, Consistent, and Highly Available Datastore

Jun Rao‡, Eugene J. Shekita†, Sandeep Tata†
†IBM Almaden Research Center
‡LinkedIn Corporation*

## ABSTRACT

Spinnaker is an experimental datastore that is designed to run on a large cluster of commodity servers in a single datacenter. It features key-based range partitioning, 3-way replication, and a transactional get-put API with the option to choose either strong or timeline consistency on reads. This paper describes Spinnaker's Paxos-based replication protocol. The use of Paxos ensures that a data partition in Spinnaker will be available for reads and writes as long a majority of its replicas are alive. Unlike traditional master-slave replication, this is true regardless of the failure sequence that occurs. We show that Paxos replication can be competitive with alternatives that provide weaker consistency guarantees. Compared to an eventually consistent datastore, we show that Spinnaker can be as fast or even faster on reads and only 5% to 10% slower on writes.

## 1. INTRODUCTION

On transactional workloads, many internet and cloud computing applications have scaling requirements that exceed the capabilities of enterprise databases. One cost effective way to meet these scaling requirements is to use *sharding* on a cluster of commodity servers. With sharding, each node of the cluster is responsible for part of the data and runs its own independent instance of the database software. To ensure linear scaling, the scope of a transaction is usually limited to a single node.

Sharding a database is often a manual process which can lead to maintenance and load balancing headaches. Recently, new partitioned database architectures have emerged [10, 12, 11, 7] that, among other things, automate the sharding and load balancing process. These new architectures typically use key-based hash or range partitioning to assign data to nodes in the cluster.

In addition to aggressive scaling requirements, many internet and cloud computing applications also need to be continuously available. However, on a large cluster of commodity servers with hundreds or even thousands of nodes, failures are inevitable. Consequently, some sort of replication strategy is needed for high availability and fault tolerance. One solution is to use synchronous master-slave replication on pairs of nodes to provide this capability. Indeed, many sharded databases are set up this way. However, master-slave replication is not an ideal solution at internet or cloud scale, as discussed below.

### 1.1 Limitations of Master-Slave Replication and the Case for Paxos

In traditional 2-way synchronous replication, one node in a master-slave pair is designated as the master and all writes are routed to it. The master's log is shipped to the slave and the master forces a commit record to disk only after the slave forces it first. If the slave goes down, the master simply continues on without the slave. Conversely, if the master goes down, the slave has the latest database state, so it can take over. Used this way, master-slave replication can normally tolerate one node going down, with the database still available for reads and writes. However, it is well known that failure sequences can be constructed where the database becomes unavailable even with just one node down.

Consider the example shown in Figure 1. The log sequence number (LSN) of the last committed write on disk is shown. Both nodes start with LSN=10 (a), then the slave goes down (b). The master continues accepting writes up to LSN=20 but then also goes down (c). Next, the slave comes back up with the master still down (d). However, at this point, the slave cannot accept reads or writes since it does not have the latest database state. Moreover, if the master suffers a permanent failure, committed writes with LSN=11 to LSN=20 are lost. The only way to avoid these problems is to block writes whenever one node in a master-slave pair goes down, but limiting availability this way may not be acceptable for some applications. While it is tempting to dismiss this example as highly improbable, in large clusters, events that are normally rare become more likely.

Double-disk failures also become more likely in large clusters [5]. With 2-way replication and no RAID or other special hardware, these can lead to catastrophic data loss. As a result, 3-way replication is commonly used with commodity servers [10, 12, 11, 7]. In addition to protecting against disk failures, 3-way replication also simplifies certain management tasks. For example, with 3 replicas, a single node failure no longer results in "panic mode", where one more failure could cause data to be lost. Online upgrades also become easier, since one replica can be taken off line and upgraded, while the other 2 replicas are kept online [7].

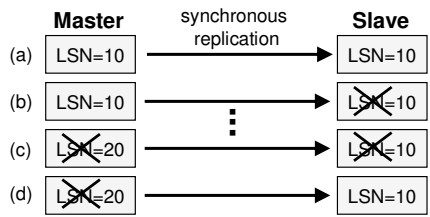

**Figure 1: Example in master-slave replication where the database becomes unavailable for reads and writes with just one node down.**

Unfortunately, with 3 replicas, more complicated failure sequences than the one shown in Figure 1 become possible. Maintaining consistency among replicas in the face of arbitrary failures has been studied in the distributed systems community for almost three decades [23]. Naive solutions often work for simple cases but have not been shown to be correct in general. The Paxos family of protocols [19, 20] is widely considered to be the only proven solution when there are 3 or more replicas. Paxos solves the general problem of reaching consensus on the state of $2F+1$ replicas while tolerating up to $F$ failures. However, Paxos has not been used for database replication because it is generally perceived as too complex and slow.

## 1.2 Strong vs. Eventual Consistency

In distributed systems, the *consistency model* describes how different replicas are kept in sync. *Strong consistency* guarantees that all replicas appear identical to applications. Although strong consistency is clearly a desirable property for building applications, it is impossible to achieve without sacrificing either availability or tolerance to network partitions. This was first observed by Brewer in his well-known CAP Theorem [6], which states that among **C**onsistency, **A**vailability, and **P**artition tolerance, only two out of three are possible.

Systems like Dynamo [12] use *eventual consistency* to provide high availability and partition tolerance for cross-datacenter replication. In CAP terminology, Dynamo is an example of an AP system. With eventual consistency, failures, network partitions, or conflicting writes can cause replicas to diverge, and applications may see multiple versions of the same data item. As a result, applications must be prepared to do conflict detection and resolution themselves. The familiar isolation guarantees of ACID transactions are not supported.

While a small class of applications with extreme availability requirements may be able to tolerate the nuances of eventual consistency, we would argue that most applications will desire stronger consistency guarantees and some support for transactions. Stonebraker has convincingly argued [2] that within a single datacenter where network partitions are rare, opting for strong consistency and availability is a better design choice, that is, picking CA in CAP.

## 1.3 Spinnaker

This paper describes a solution to the consistent replication problem in the context of Spinnaker, which is an experimental datastore that is designed to run on a large cluster of commodity servers in a single datacenter. We use the term "datastore" to distinguish Spinnaker from a full fledged relational database. Spinnaker features key-based range partitioning, 3-way replication, and a transactional get-put API with the option to choose either strong or *timeline consistency* [11] on reads. Timeline consistency allows potentially stale data to be returned in exchange for better performance.

For replication, Spinnaker uses a Paxos-based protocol that is integrated with its commit log and recovery processing. The use of Paxos ensures that a data partition will be available for reads and writes as long as a majority of the nodes containing its replicas are alive. In CAP terminology, Spinnaker is an example of a CA system. It is designed for a single datacenter and assumes that a different replication strategy is used for cross-datacenter fault tolerance (presumably an asynchronous one).

The remainder of this paper presents an overview of Spinnaker's data model, API, and architecture, followed by a detailed description and evaluation of its replication protocol. The main contributions of the paper are as follows: We show how Paxos replication can be integrated with the logging and recovery of a scalable datastore. We show that a Paxos implementation can be simpler than previously assumed through careful design choices and the use of a distributed coordination service. We also show that the performance of Paxos replication can be competitive with other alternatives that provide weaker consistency guarantees. Compared to an eventually consistent datastore, we show that Spinnaker can be as fast or even faster on reads and only 5% to 10% slower on writes. Finally, we show that Spinnaker's design leads to a highly available system, with node recovery taking less than half a second.

## 2. RELATED WORK

### 2.1 Two-Phase Commit

Two-phase commit (2PC) has been suggested as a way to maintain consistent replicas [26]. However, 2PC treats each participant as an independent resource manager with its own database (versus just a replica). Consequently, for replication, 2PC is overkill and suffers from several disadvantages. First, the failure of a single node will lead to an abort. This is unacceptable when the aim is to keep the system available in the presence of node failures. Second, invoking 2PC for every transaction will lead to poor performance, since a typical implementation of 2PC requires 2 disk forces and 2 message delays. Finally, 2PC blocks when the coordinator fails. Non-blocking three-phase commit protocols [25] have been proposed but are seldom used in practice because of their poor performance.

### 2.2 Database Replication

There is a substantial body of work on database replication. In contrast to Spinnaker, work in this area has focused on replication in the context of a single unpartitioned database. Postgres-R [18] was one of the earlier systems to be described. It uses a group communication system (GCS) to order and replicate transactions using reliable multicast. While a GCS can be built to use a consensus protocol like Paxos, such an approach is not likely to perform as well as the one taken in Spinnaker, where the replication protocol is tightly integrated with the commit log and recovery processing. The designers of Tashkent [14] made a similar observation and suggested co-locating transaction ordering and

244

logging to improve performance. Their solution requires a "certifier" to coordinate replication. However, they did not specify how the certifier itself could be replicated and made fault tolerant.

Cecchet et al. [8] provides a good overview of middleware-based solutions for replication. Ganymed [24] is an example of a middleware solution that uses a single master and FIFO queues to replicate data. While the middleware solutions that have been described in the literature are able to recover from simple failure scenarios, it is unclear they can handle some of the more complicated failure scenarios that Paxos was designed to handle.

## 2.3 Dynamo, Bigtable, and PNUTS

Amazon's Dynamo [12], which was mentioned earlier, is a scalable key-value store that uses eventual consistency to provide high availability and partition tolerance. Conflicts caused by eventual consistency are resolved using vector clocks. To try and keep replicas in sync, background "anti-entropy" measures like "read-repair" and "merkle trees" are used. Spinnaker's architecture is arguably simpler than Dynamo's because there is no need for conflict resolution and only one mechanism, namely Paxos, is used to keep replicas in sync.

Google's Bigtable [10] is a scalable datastore that provides strong consistency and support for single-operation transactions. In contrast to Spinnaker, each Bigtable node relies on a distributed file system (GFS) [15] for storing its data and log, as well as for replication. Relying on GFS simplifies Bigtable, but has several drawbacks for transactional workloads. For example, forcing a log page to disk incurs a fair bit of overhead, requiring communication with a centralized GFS master and remote copies of the page to be forced and acknowledged. There is also no notion of a hot standby. When a Bigtable node goes down, all the data on that node becomes unavailable until the node is restarted and its log in GFS is replayed. In a recent interview, Google engineers conceded that GFS is a poor fit for transactional workloads and are reported to be working on an alternative design [22].

Yahoo's PNUTS [11] is scalable datastore that supports timeline consistency and single-operation transactions. In contrast to Spinnaker, PNUTS is more focused on the problem of cross-datacenter replication. PNUTS relies on a centralized pub-sub service called the Yahoo Message Broker (YMB) for replication. YMB is critical to the performance of PNUTS, but unfortunately the details of YMB have not been made public.

## 2.4 Other Systems

Microsoft's SQL Azure, which provides a cloud SQL service, was described in [7]. Although the details of SQL Azure's replication strategy have not been described, we speculate that the techniques used in Spinnaker could be applied to it as well. FAWN [4] is a scalable key-value store with more of a focus on low-power servers. Finally, the conditional put call in Spinnaker, which is discussed in Section 3, is very similar to the notion of a minitransaction in Sinfonia [3].

## 3. DATA MODEL AND API

Spinnaker's data model and API are similar to those in Bigtable and PNUTS. Data is organized into rows and tables, with each row in a table uniquely identified by its key. A row may contain any number of columns with corresponding values and version numbers. Column names and values are opaque bytes to Spinnaker. The basic API is as follows:

**get(key, colname, consistent)**
Read a column value and its version number from a row. The setting of the 'consistent' flag is used to choose the consistency level. Setting it to 'true' chooses strong consistency, and the latest value is always returned. Setting it to 'false' chooses timeline consistency, and a possibly stale value is returned in exchange for better performance.

**put(key, colname, colvalue)**
Insert a column value into a row.

**delete(key, colname)**
Delete a column from a row.

**conditionalPut(key, colname, value, v)**
Insert a new column value into a row only if the column's current version number is equal to 'v'. Otherwise, an error is returned.

**conditionalDelete(key, colname, v)**
Like conditional put but for delete.

Version numbers are monotonically increasing integers that are managed by Spinnaker and exposed through its get API. These are used in conditional put and delete to provide a simple form of optimistic concurrency control for read-modify-write transactions on a row. For example, to transactionally increment a counter $c$, an application would use:

c = get(key, "c", consistent=true);
ret = conditionalPut(key, "c", c.value + 1, c.version);

In this example, the application is assumed to have logic in place to retry the increment if an error is returned.

Note that each API call is executed as a single-operation transaction. Although it has not been shown, Spinnaker also provides multi-column versions of its API. For example, the multi-column version of conditional put allows multiple columns of the same row to be conditionally put with one API call.

For ease of explanation, the remainder of this paper will denote any API call that does not modify data as a *read* and any API call that modifies data as a *write*.

## 4. ARCHITECTURE

This section presents a brief overview of Spinnaker's architecture. Space limitations prevent a detailed discussion of issues such as logical to physical node assignment, adding and dropping nodes, load balancing, and so on. The goal here is to provide just enough background to understand Spinnaker's replication protocol, which is the main focus of this paper.

Like Bigtable and PNUTS, Spinnaker distributes the rows of a table across its cluster using range partitioning. Figure 2 shows an example of a Spinnaker cluster with 5 nodes. Each node is assigned a base key range, which is replicated on the next $N - 1$ nodes ($N = 3$ by default). For example, in Figure 2, node A's base key range is [0, 199], which is replicated on nodes B and C. This style of replication is similar to chained declustering [16]. We will always assume the default replication setting of $N = 3$ in this paper.

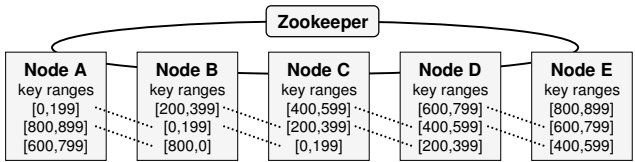

**Figure 2: Example of a Spinnaker cluster.**

Each group of nodes involved in replicating a key range is denoted as a *cohort*. Note that cohorts overlap. For example, in Figure 2, nodes A-B-C form the cohort for key range [0, 199], nodes B-C-D form the cohort for key range [200, 399], and so on.

### 4.1 Node Architecture

Each node in Spinnaker contains several components, as shown in Figure 3. All the components are thread safe, allowing multiple threads to support each of the 3 key ranges on a node. A shared write-ahead log allows a dedicated logging device to be used for performance. Each log record is uniquely identified by an LSN. In order to share the same log, each cohort on a node uses its own logical LSNs. The *commit queue* is a main-memory data structure that is used to track pending writes. Writes are committed only after receiving a sufficient number of acks from a cohort. In the meantime, they are stored in the commit queue.

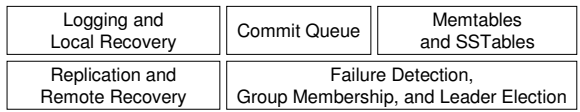

**Figure 3: The main components of a node.**

Committed writes are placed in a *memtable*, which is periodically sorted and flushed to an immutable disk structure called an *SSTable*. SSTables are indexed by key and column name for efficient access. In the background, smaller SSTables are merged into a larger ones to garbage collect deleted rows and improve read performance. Spinnaker's SSTables are based on the design used in Bigtable. Further details are beyond the scope of this paper and the reader is referred to [10].

### 4.2 Zookeeper

Zookeeper [17] is used as a fault tolerant, distributed coordination service in Spinnaker. By providing a centralized place to store meta-data and manage events like node failures, Zookeeper greatly simplifies Spinnaker's design. The combination of primitives supported by Zookeeper make it fairly easy to implement distributed locks, barriers, group membership, and so on.

More will be said about Zookeeper later. It is important to note that Zookeeper is *not* in the critical path of reads and writes in Spinnaker. Normally, the *only* messages exchanged between a Spinnaker node and Zookeeper are heartbeats. As a result, we expect a single Zookeeper service to support a Spinnaker cluster with thousands of nodes.

## 5. THE REPLICATION PROTOCOL

This section describes Spinnaker's replication protocol, which runs on a per-cohort basis. Some familiarity with

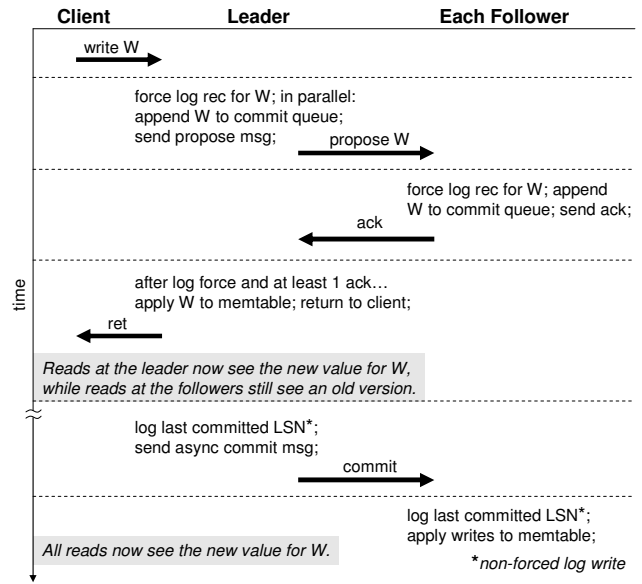

**Figure 4: The replication protocol.**

Paxos is helpful to understand this section. See Appendix A for a brief overview of Paxos and a discussion of how Spinnaker's Paxos implementation differs from a traditional one.

Each cohort has an elected *leader*, with the other 2 nodes acting as *followers*. The replication protocol has two phases: a *leader election* phase, followed by a *quorum phase* where the leader *proposes* a write and the followers *accept* it. The leader election phase is described in Section 7. In the absence of failures, the leader does not change, and only the quorum phase needs to be executed.

Figure 4 shows the flow of Spinnaker's replication protocol in steady state. When a client submits a write $W$, it always gets routed to the leader of the affected key range. The leader appends a log record for $W$ to its log and then initiates a log force to disk. In parallel with the log force, the leader appends $W$ to its commit queue and sends a *propose message* for $W$ to its followers.

When the followers receive the propose message, they force a log record for $W$ to disk, append $W$ to their commit queue, and reply with an ack to the leader. After the leader gets an ack from at least one follower, it applies $W$ to its memtable, effectively committing $W$. Finally, the leader returns a response to the client. Note that, unlike traditional write-ahead logging, there is no separate commit record for $W$ that needs to be logged. This is because $W$ is performed as a single-operation transaction. The quorum-based recovery procedure, which is described in the next section, ensures the durability of $W$ through re-proposals.

Periodically, the leader sends an asynchronous *commit message* to the followers asking them to apply all pending writes up to a certain LSN to their memtable, effectively committing those writes. For recovery, the leader and followers also save this LSN, which is referred to as the *last committed LSN*, using a non-forced log write.

Note that strongly consistent reads are always routed to the cohort's leader, so they are guaranteed to see the latest value for $W$. However, timeline consistent reads can be routed to any node in the cohort, so they may see a stale value for $W$ until its commit message is processed. The interval for commit messages is called the *commit period*.

The staleness of followers can be reduced by decreasing the commit period.

In total, 3 log forces and 4 messages are needed to commit a write. However, many of these operations are overlapped. Although it has not been discussed, group commit [13] is also used to improve logging performance. In terms of latency, the critical path is the time for a follower to get the propose message, force its log, and send an ack to the leader, i.e., 1 log force and 2 message delays.

## 5.1 Conditional Put

Replication, logging, and recovery for conditional put are the same as with the regular put call. The only difference is that, before executing a conditional put, the cohort's leader checks to see if the current version of the column being written matches the version specified in the call. If so, the write is executed. Otherwise, no data is written and an error code is returned to the client. Note that a conditional put is guaranteed to have the same outcome on each node of the cohort because writes are executed in LSN order within a cohort.

## 6. RECOVERY

This section describes how a cohort's leader and followers are recovered after a node failure. To simplify the discussion, these are described from the perspective of a single cohort. In practice, the 3 cohorts that a node belongs to are actually recovered in parallel with a shared scan of the node's log.

## 6.1 Follower Recovery

The recovery of a follower proceeds in two phases: *local recovery* and *catch up*. Let $f.cmt$ and $f.lst$ denote the follower's last committed LSN and the last LSN in its log, respectively, where $f.cmt \leq f.lst$. In the local recovery phase, the follower can safely re-apply log records from its most recent checkpoint thru $f.cmt$ to recover the state of its memtable, and this is done in an idempotent way. However, the state of writes after $f.cmt$ are ambiguous – they may or may not have been committed by the cohort's leader. The state of these log records is resolved during the catch up phase. If the follower has lost all its data because of a disk failure, then it moves directly to the catch up phase.

In the catch up phase, the follower advertises $f.cmt$ to the leader. The leader responds by sending all committed writes after $f.cmt$ to the follower. At the end of the catch up phase, the leader momentarily blocks new writes to ensure that that the follower is fully caught up.

In practice, the oldest part of a node's log is rolled over when its writes have been captured to an SSTable. As a result, the leader may no longer have access to some of the log records it needs in the catch up phase. To work around this, each SSTable is tagged with the min and max LSN of the writes that it contains. When a catch up request cannot be served from the leader's log, the appropriate SSTable is located and sent to the follower.

### 6.1.1 Logical Truncation of the Follower's Log

As mentioned, the state of writes after $f.cmt$ are ambiguous and may not have been committed. It is possible that the leader went down, a new leader was elected, and the new leader discarded some of the log records after $f.cmt$. These discarded log records need to be removed from the follower's log to ensure that they are never re-applied by future invocations of local recovery. At first glance, it might seem like the follower's log could just be truncated to $f.cmt$ to solve this problem. However, this is not possible because the follower's log is shared by multiple key ranges (i.e., different cohorts), and some of the log records following $f.cmt$ may be needed to recover another cohort.

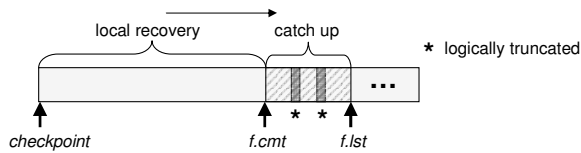

Figure 5: Logical log truncation.

The solution to this problem is *logical truncation* of the follower's log. The LSNs of log records belonging to the follower between $f.cmt$ and $f.lst$ are remembered in a *skipped-LSN* list, which is saved to to a known location on disk. Future invocations of local recovery will check the skipped-LSN list before processing log records. Since this list is expected to be small, it is loaded into memory before recovery. Skipped-LSN lists are managed and garbage-collected along with log files.

Figure 5 shows the parts of the follower's log that are involved in local recovery, the catch up phase, and logical truncation. For a more detailed example, see Appendix B.

## 6.2 Leader Takeover

A cohort's key range becomes unavailable for writes when its leader fails. When the leader fails, a new leader is elected and *leader takeover* occurs. The new leader is chosen in a way that ensures its log will contain every write committed by the old leader. More will be said about this in Section 7. Note that if the old leader subsequently comes back up, it will rejoin the cohort as a follower and run the follower recovery procedure just described.

When the old leader failed, it may have had some writes that it committed but were still in the pending state at the followers because a commit message for them had not been sent yet. The new leader finishes committing these "unresolved" writes using the algorithm in Figure 6.

---

1. let $l.cmt$ = the leader's last committed LSN;
2. let $l.lst$ = the leader's last LSN;
3. **for** (each follower $f$) **do**
4.     let $f.cmt = f$'s last committed LSN;
5.     send committed writes in $(f.cmt, l.cmt]$ to $f$;
6.     send a commit msg with $l.cmt$ to $f$;
7. **end for**
8. wait until at least one follower is caught up to $l.cmt$;
9. re-propose writes in $(l.cmt, l.lst]$ to followers, commit these using the normal replication protocol;
10. open the cohort for writes;

---

Figure 6: Leader takeover.

As shown, leader takeover begins by catching up each follower to the new leader's last committed LSN (lines 1-7). If a follower has failed and is also in recovery, this corresponds to the follower catch up phase described in the previous section, but from the leader's perspective. Otherwise, a follower may already have some of the writes that are sent by the new leader (line 5). By checking LSNs, these can be detected and ignored by the follower.

Using the notation in Figure 6, the writes between $l.cmt$ and $l.lst$ are the ones the new leader needs to resolve. After the new leader has a quorum (line 8), these unresolved writes are re-proposed (line 9) and committed using the normal replication protocol. Finally, the leader opens the cohort for writes (line 10), starting with an LSN that is larger than any LSN previously used in the cohort.[1]

## 7. LEADER ELECTION

This section describes Spinnaker's leader election protocol, which is run on a per-cohort basis. Leader election is triggered whenever a cohort's leader has failed or following local recovery after a system restart. The leader election protocol has to guarantee that the cohort will reach a consensus and that the new leader is chosen in a way that no committed writes are lost. Implementing a distributed protocol with these guarantees in the presence of arbitrary failures is non-trivial, but Zookeeper greatly simplifies the task. Before we can describe how leader election works, a brief overview of Zookeeper's data model and API is required.

### 7.1 Zookeeper's Data Model and API

Zookeeper's data model resembles a directory tree in a file system, with a node in the tree identified by its path from the root, for example, /a/b/c. Each so-called *znode* includes associated binary data. A client can create and delete znodes in a directory, with Zookeeper taking care of serializing changes and making sure they are reliably stored on disk.

A znode can be either *persistent* or *ephemeral*. Zookeeper automatically deletes an ephemeral znode if the client that created it dies, whereas a persistent znode has to be explicitly deleted. A znode can also include a *sequential* attribute, causing Zookeeper to add a unique, monotonically increasing identifier to the znode when it is created. For event handling, a client can set a *watch* on a znode. These cause the client to be notified of any changes to the znode or its children.

### 7.2 The Leader Election Protocol

Each Spinnaker node includes a Zookeeper client that, among other things, is used to run the leader election protocol shown in Figure 7. Only a sketch of the protocol can be presented here, with certain race conditions ignored, but readers should still be able to get a reasonable understanding of how the protocol works.

Let $r$ correspond to the key range of the cohort running leader election. Information needed for leader election is stored in Zookeeper under /r. Leader election begins with one of the cohort's nodes cleaning up any state from a previous round of leader election (line 1). Next, each node of the cohort announces itself as a candidate in the election by advertising its last LSN ($n.lst$) in a sequential ephemeral znode under /r/candidates (line 4). Then a watch is set on /r/candidates, causing Zookeeper to notify the cohort whenever /r/candidates changes (line 5).

Once a majority of the cohort (i.e., 2 nodes) appears under /r/candidates, each node in the cohort checks to see if

---
[1] In practice, this is implemented by reserving the high order bits of an LSN for an *epoch* number, which is saved in Zookeeper and incremented every time a new leader takes over. For a detailed example, see Appendix B.

---

1. clean up old state under /r if necessary;
2. let $n$ = this node;
3. let $n.lst$ = this node's last LSN;
4. add a sequential ephemeral znode to /r/candidates with value = $n.lst$;
5. set a watch on /r/candidates and wait for a majority;
6. the new leader is the candidate with the max $n.lst$, using znode sequence numbers to break ties;
7. **if** ($n$ is the leader and /r/leader is empty) **then**
8.     write an ephemeral znode under /r/leader with value = $n$.hostname;
9.     execute leader takeover;
10. **else**
11.     read /r/leader to learn the new leader;
12. **end**

---

**Figure 7: Leader election.**

it is the new leader. The new leader is the candidate with the max $n.lst$ (line 6), using Zookeeper sequence numbers to break ties. Next, the new leader writes its hostname in /r/leader and runs the leader takeover algorithm described earlier (lines 8-9). An ephemeral znode is used in this step so the cohort can be notified if the new leader subsequently fails. Finally, followers learn about the new leader by reading /r/leader (line 11).

As noted above, the leader election protocol has to guarantee that the cohort will reach a consensus and that the new leader is chosen in a way that no committed writes are lost. It should be clear that Figure 7 reaches a consensus on the new leader. Proving that no committed writes are lost is a little more subtle. This stems from the fact that a committed write has to be forced to the logs of at least 2 nodes in the cohort, and at least 2 nodes in the cohort have to participate in leader election. With 3 nodes in the cohort, this means that at least one of the nodes participating in leader election will have the last committed write in its log. By choosing the node with the max $n.lst$ (line 6), the protocol ensures that the new leader will have this write in its log. And if the write is still in an unresolved state on other nodes in the cohort, leader takeover will make sure it is re-proposed.

Each Spinnaker node also includes an event handler, which runs as a Zookeeper client. We omit the details of how the event handler works, but it basically interacts with Zookeeper to coordinate things like node failures, a failed node coming back up and rejoining its cohort, and so on. The event handler is what calls the leader election protocol just described.

## 8. DISCUSSION

### 8.1 Availability and Durability Guarantees

With the default replication setting of $N = 3$, Spinnaker will not commit a write until it has been forced to the logs of 2 out of 3 nodes in a cohort. A cohort continues to be available for strongly consistent reads and writes as long as a majority (2) of its nodes are up. Unlike traditional master-slave replication, this is true regardless of the failure sequence that occurs. With timeline consistency, a cohort continues to be available for reads even if just 1 node in the cohort is up.

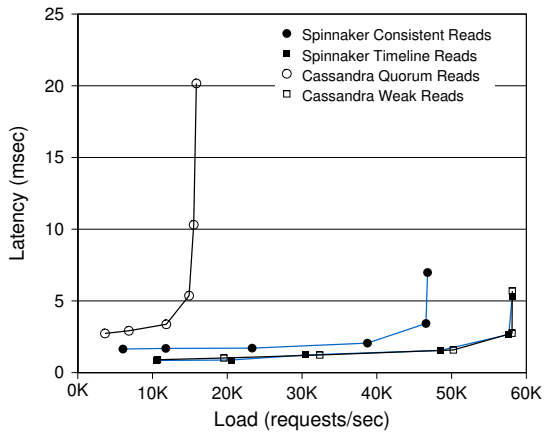

Figure 8: Average read latency.

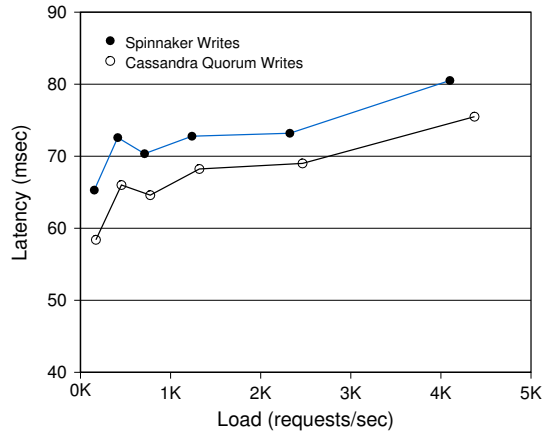

Figure 9: Average write latency.

Under normal circumstances, a cohort will not lose committed data even if 2 out of 3 of its nodes permanently fail. However, a small window of committed writes can be lost if a cohort's leader and one of its followers permanently fail in rapid succession.

### 8.2 Multi-Operation Transactions

Currently, each API call in Spinnaker is executed as a single-operation transaction. However, we believe that multi-operation transactions could be supported with fairly modest extensions to its replication protocol and recovery procedures. This is a promising direction for future work that, in addition to enhancing Spinnaker, could also be applied to a scalable SQL cluster, with full-fledged relational nodes. The basic idea would be to let a transaction create multiple log records, but only invoke the replication protocol for a batch of log records at commit time. Then during recovery, the replicas of the log would first be brought into a consistent state using Paxos, followed by a local (per-node) redo and undo recovery pass.

### 8.3 Design Tradeoffs

Spinnaker represents just one solution in the complex space of design choices for a scalable datastore. On the positive side, Spinnaker is able to provide strong or timeline consistency, which we believe makes it appealing to a wider range of applications than an eventually consistent datastore. By using Paxos replication on a per-cohort basis, Spinnaker is able to do this in a scalable and highly available manner. On the negative side, all the writes for a cohort have to be routed to the cohort's leader in Spinnaker. If strong consistency is chosen, all the reads for a cohort also have to be routed to the cohort's leader. Consequently, Spinnaker's performance, scaling, and availability can be lower than an eventually consistent datastore like Dynamo. The experimental results in the next section and Appendix D were designed to examine these tradeoffs.

### 9. EXPERIMENTAL RESULTS

This section presents the results of experiments that were run to compare Spinnaker's performance with an eventually consistent datastore, namely Cassandra [1], which is an open-source datastore based on Dynamo. For details on the setup used for experiments see Appendix C.

Like Dynamo, Cassandra provides knobs for tuning its consistency, availability, and durability. Spinnaker is actually derived from the Cassandra codebase, making for a fair comparison. Both datastores share a similar data model and support 3-way replication.

Among other things, Cassandra includes support for *weak reads* and *quorum reads*. A weak read accesses just 1 replica, whereas a quorum read accesses 2 replicas to check for conflicts caused by eventual consistency. Conflicts are resolved using timestamps. Cassandra also includes support for *weak writes* and *quorum writes*. Both are sent to all 3 replicas, but a weak write waits for an ack from just 1 replica, whereas a quorum write waits for acks from 2 replicas.

The authors of Dynamo recommended using quorum reads and writes to minimize conflicts caused by eventual consistency [12], Note, however, that quorum reads and writes still do not guarantee strong consistency in Dynamo (or Cassandra). This is because there is no notion of a cohort leader to serialize writes, so conflicts can still occur if there are concurrent writes to different replicas. The lack of a quorum-based recovery algorithm also means there is no guarantee that a replica will be brought up to a consistent state after a node failure. Therefore, in the discussion that follows, readers should bear in mind that Cassandra and Spinnaker really provide two different consistency models.

### 9.1 Read Results

In this experiment, each client read 4KB values from random rows. The performance of Spinnaker's strongly consistent and timeline consistent reads, referred to as *consistent reads* and *timeline reads*, respectively, were compared to Cassandra's weak and quorum reads.

Figure 8 shows the average latency of a read as the load was increased. As shown, the latency of Cassandra's quorum read was 1.5x to 3.0x worse than Spinnaker's consistent read. The knee of the latency curve also occurred much sooner in Cassandra. This is because a quorum read in Cassandra has to access 2 replicas and check for conflicts, whereas a consistent read in Spinnaker only has to access the cohort leader's replica.

Not surprisingly, Cassandra's weak read had the best latency. However, many applications may not be able to live with the consistency guarantees that it provides, which can be unpredictable. Spinnaker's timeline read is probably a more practical choice if slightly stale values can be toler-ated. As shown, its latency was nearly identical to Cassandra's weak read.

## 9.2 Write Results

In this experiment, each client wrote 4KB values into rows with consecutive keys. On Cassandra, only quorum writes were considered, since they provide the same durability guarantees as Spinnaker, returning only after a write has been logged to 2 disks.

Figure 9 shows the average latency of a write as the load was increased. As shown, the latency of Spinnaker's write was 5% to 10% worse than Cassandra's quorum write across the board. We believe that this is a small price to pay for strong consistency. The added latency is because a write in Spinnaker has to wait for an ack from a cohort leader and one of its followers, whereas a quorum write in Cassandra only has to wait for acks from *any* 2 replicas. When the cohort leader gets heavily loaded, this can hurt Spinnaker's write latency.

Note that the write latency of both Spinnaker and Cassandra is rather poor compared to a commercial database. This is mainly because Cassandra's log manager, which was reused in Spinnaker, is fairly primitive and can incur unwanted disk seeks. However, we found that storing the log on a solid-state disk (SSD) eliminated this problem, with the average write latency improving to a respectable 6 msec or less in most cases for both Spinnaker and Cassandra. See Appendix D for more details.

## 10. CONCLUSION

This paper described the Paxos-based replication protocol used in Spinnaker, which is a scalable, consistent, and highly available datastore that is designed to run on a large cluster of commodity servers in a single datacenter. The use of Paxos and 3-way replication ensures that a data partition in Spinnaker will be available for reads and writes as long a majority of its replicas are alive. Unlike traditional master-slave replication, this is true regardless of the failure sequence that occurs.

Paxos has not been used for database replication because it is generally perceived as too complex and slow. However, we showed that a distributed coordination service like Zookeeper makes it significantly easier to implement Paxos, and that the performance of Paxos can be acceptable within a single datacenter. Compared to Cassandra, which is an eventually consistent datastore, we showed that Spinnaker is as fast or even faster on reads and only 5% to 10% slower on writes.

In terms of future work, many aspects of Spinnaker are open to further study. These include adding support for multi-operation transactions, online algorithms for load balancing that work with its replication protocol, and a detailed comparison to a datastore like Bigtable, which relies on DFS-based replication.

# APPENDIX
## A. PAXOS OVERVIEW

Getting a group of nodes to agree on a given value is described as the *consensus* problem [23] in the distributed systems literature. This is known to be a difficult problem when nodes can fail and messages can be lost or delivered out of order. The Paxos family of protocols [19, 20] solves the general problem of reaching consensus on the state of $2F+1$ replicas while tolerating up to $F$ failures. Once consensus has been reached, Paxos guarantees that the value can be retrieved from the group as long as a majority of its nodes are up. To reach consensus on a value for a data item $D$, the protocol proceeds in two phases:

**1a. Propose:** A node wishing to propose its value $v$ for $D$ chooses a proposal number $n$ and sends a *prepare* message to the nodes in the group.

**1b. Promise:** Suppose a node receives a prepare message with a proposal number $n$. If $n$ is greater than any previously accepted prepare message, the node responds with a *promise* message, promising not to accept any new proposals numbered less than $n$. Otherwise, the node responds with a *nack*. If the node has accepted a lower numbered proposal, the promise message includes the value accepted.

**2a. Accept:** If a proposer receives promises from a majority of the group, it sends an *accept* message to these nodes specifying $v$ and $n$. If any of the nodes responded with a previously accepted value, then the proposer is required to pick the value with the largest $n$ among the responses.

**2b. Ok:** A node that receives an accept message with value $v$ and proposal number $n$ responds by accepting $v$ and sending an *ok* message. However, if the node has already accepted a prepare message with a proposal number greater than $n$, then no response is given.

Phase 1, consisting of the propose and promise steps, effectively elects a leader for the group. In Phase 2, the group leader sends the value to be accepted and the rest of the nodes accept it with the ok message. This is roughly equivalent to a quorum phase.

Nodes are required to choose unique, monotonically increasing proposal numbers for the protocol to work correctly. Also, nodes write their actions to stable storage in a write-ahead log before sending messages to other nodes. This is a simplified description of Paxos and we encourage the reader to look at [19, 20, 9] for more background.

Paxos solves the problem of getting a group of nodes to agree on a single value. *Multi-Paxos* is a well known optimization of Paxos when a sequence of values are being submitted to the group. Assuming the leader is relatively stable, Multi-Paxos skips leader election and simply executes the quorum phase. With this optimization, a value submitted to the leader can be acknowledged after 2 message delays, with an accept and ok message.

## A.1 Spinnaker's Variation on Multi-Paxos

Spinnaker's replication protocol is based on a variation of Multi-Paxos. Some of the modifications that were made to the basic Multi-Paxos protocol are worth noting here. Other implementations of Multi-Paxos, such as those in Zookeeper and Chubby [9], have included similar modifications.

One modification is the way Spinnaker integrates its replication, database commit processing, and recovery within the same protocol framework. For efficiency, a shared write-ahead log is used for all of these. Another modification is Spinnaker's catch up phase, which is run during recovery to ensure that a node is not missing any log entries. In contrast, basic Multi-Paxos allows a node that has failed to come back and participate in the next round of Paxos right away. This would be unacceptable in Spinnaker, since applying a log with missing entries would introduce data inconsistencies.

Spinnaker uses reliable in-order messages based on TCP sockets to simplify its replication protocol. This is a practical design choice that was also made in Zookeeper [17]. In contrast, Multi-Paxos assumes an unreliable message layer.

Finally, Spinnaker relies on a distributed coordination service, namely Zookeeper, for leader election. As noted above, Zookeeper is also based on Paxos. The idea of using a coordination service, itself based on Paxos, to simplify and scale-out other Paxos implementations (in this case Spinnaker), was also advocated by Lamport et al. in their paper on "Vertical Paxos" [21].

## B. RECOVERY EXAMPLE

In this section, we present a detailed recovery example where the leader and the followers of a cohort simultaneously fail. We walk through the different steps during recovery to illustrate leader takeover, follower catch up, and logical log truncation.

LSNs are denoted using a two-part representation $e.seq$, where $e$ is the *epoch* number and $seq$ is the sequence number. In practice, the high order bits of the LSN are used to store the epoch number, while the low order bits are used to store the sequence number. Epoch numbers are Spinnaker's mechanism to guarantee that on leader takeover new writes are assigned LSNs that are greater than any previously used LSN in the cohort. As part of leader takeover, a new epoch number is stored in Zookeeper before the leader accepts any new writes. Readers familiar with Paxos will note that LSNs effectively play the role of proposal numbers.

Consider the example in Figure 10. The cohort consists of nodes A, B, and C. Each node is shown with its last committed LSN ($cmt$) and last LSN ($lst$). The initial state of the cohort is S0, with node A as the leader. The write for LSN 1.20 has been committed, but nodes B and C have not received an asynchronous commit message for it yet, while the writes for LSNs 1.21 and 1.22 have been proposed but not committed. Recall that writes are proposed to the cohort in parallel. This is how the writes for LSNs 1.21 and 1.22 are able to appear in the logs of the followers (nodes B and C) before they appear in the log of the leader (node A).

Next, suppose all the nodes go down, resulting in state S1, and then nodes A and B come back up, causing recovery to transition from S1 to S2. Node B is elected as the leader, since it has the largest $lst$ value. LSN 1.22 is not seen and is effectively discarded because node C is still down. This is ok, since its write was never committed. During leader takeover, node B re-proposes and commits writes for LSNs 1.11 to 1.21. Then node B increments the epoch number and begins accepting new writes from clients. Soon afterwards, writes for LSNs 2.22 to 2.30 arrive and are committed, resulting in state S3.

Finally, suppose that node C comes back up, causing recovery to transition from S3 to S4. During node C's catch up phase, node B sends it writes for LSNs 1.11 to 1.21 from epoch 1 and writes for LSNs 2.22 to 2.30 from epoch 2, with LSN 1.22 being logically truncated.

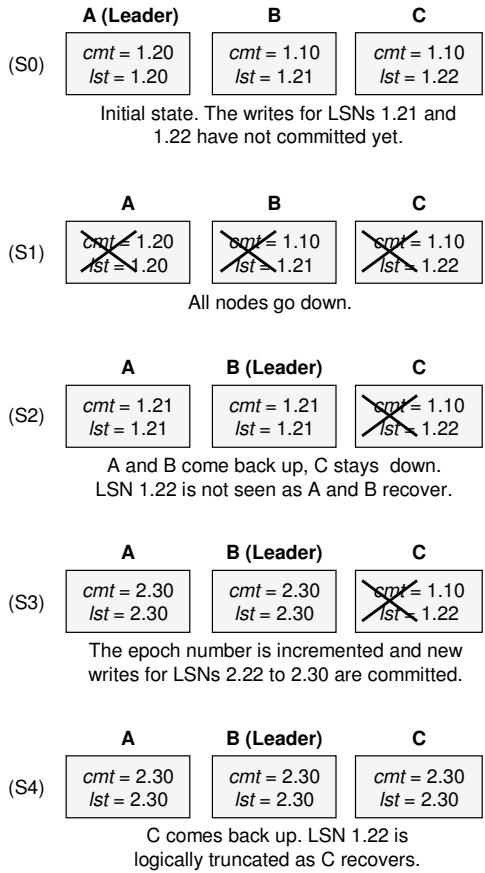

Figure 10: Detailed recovery example.

Although it is not described in the section on recovery (Section 6), a simple optimization is used in practice to decrease the number of log records that need to be sent during recovery. For example in the transition from S1 to S2, it should be clear that the new leader (node B) can re-propose just the write for LSN 1.21 to node A instead of the writes for LSNs 1.11 to 1.21. The changes required to support this optimization are straightforward.

## C. EXPERIMENTAL SETUP

Experiments were run on a cluster of 10 nodes, each with two quad-core 2.1 GHz AMD processors and 16GB of memory. Each node had 5 locally attached SATA disks, with 1 disk used as a dedicated logging device, and the remaining disks configured as a striped logical volume. To guarantee durability, we turned off the write-back cache on the SATA disks. Nodes were connected using a rack-level 1-Gbit Ethernet switch. A similar cluster with 10 nodes and its own rack-level switch was used for clients. The two clusters were connected with a second-level 1-Gbit switch.

We used the Cassandra [1] trunk as of October 2009. Spinnaker's code is based on a branch of Cassandra. Spinnaker reused Cassandra's implementation of SSTables, memtables, and the log manager, but its replication protocol, recovery algorithms, and commit queue were implemented from scratch. Zookeeper version 3.2.0 was used.

In the graphs of our results, we show the average latency of a read or write operation (the Y axis) for a given system "load" (the X axis). The latency measured included the round-trip message delay from the client to the datastore and back again to the client. For system load, we increased the number of threads per client node by powers of two and measured the average number of read or write requests per second generated by a client node. Note that the values for load varies from line to line in our graphs. This is because load is actually a function of the underlying independent variable, namely, the number of threads per client node.

In our read experiments, most of the data was cached in the memory of each datastore node, causing the CPU and network to be bottleneck in terms of latency. This is typical for most transactional workloads today. In our write experiments, the log forces required for commit processing were almost always the bottleneck. Again, this is typical for transactional workloads.

Unfortunately, Cassandra's log manager, which was reused in Spinnaker, is fairly primitive compared to the ones found in commercial databases. It includes some basic optimizations like group commit [13], but more advanced optimizations like using async I/O, multiple log buffers, and preallocated log files are missing. The lack of preallocated log files is perhaps the biggest shortcoming, since it can cause the underlying file system to update its metadata on disk as a log file grows. This in turn can cause unwanted seeks on the logging disk. We are confident that the average write latency in both Spinnaker and Cassandra could be dramatically improved with a better log manager. However, the relative difference between the two datastores would probably stay the same.

## D. ADDITIONAL EXPERIMENTS

### D.1 Availability Results

This experiment was designed to study Spinnaker's availability. A single client wrote 4KB values into rows with consecutive keys in a way that ensured all writes were routed to the same cohort leader. Then we caused the leader to fail and measured the time it took for the cohort to recover and become available for writes again.

The cohort's recovery time relative to the commit period are shown in Table 1. The results include the time for the cohort to run leader election and for the new leader to clean up any unresolved state. The 2-second timeout we used for Zookeeper to detect node failures was excluded to focus on just Spinnaker's performance.

| Commit Period (sec) | 1 | 5 | 10 | 15 |
|---|---|---|---|---|
| Recovery Time (sec) | 0.4 | 1.5 | 2.6 | 4.0 |

Table 1: Cohort recovery time.

As shown, the cohort's recovery time (i.e., unavailability window) was proportional to the commit period. Recall that the new leader needs to re-propose and commit all the log records that are in its log but were not committed by the old leader. The number of these log records is proportional to the commit period – the interval after which a leader sends an asynchronous commit message. A value of 1 second is a reasonable choice for the commit period, resulting in recovery times of less than half a second. The commit period can be made substantially smaller without much overhead by piggy-backing the commit message on propose messages for new writes.

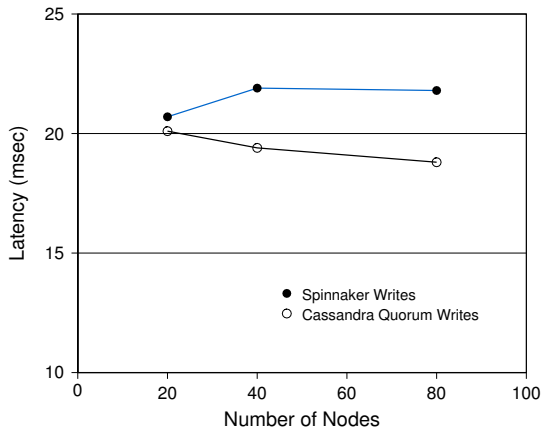

Figure 11: Average write latency with increasing cluster size on EC2.

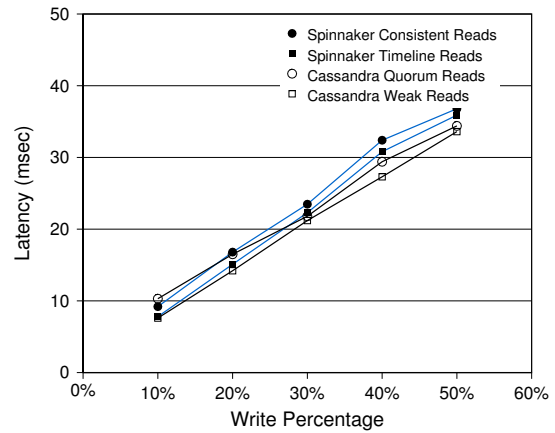

Figure 12: Average latency on a mixed workload of reads and writes.

In contrast to Spinnaker, an eventually consistent datastore like Cassandra is always available, but this is achieved by giving up consistency. Applications that can deal with brief unavailability windows can benefit from the stronger consistency model provided by Spinnaker.

## D.2 Scaling Results

To test Spinnaker's scaling, we re-ran our write experiments on Amazon's EC2 with 20, 40, and 80 standard extra-large instances. We fixed the per-node load and measured the average write latency in each case. The results in Figure 11 show that the write latency remained roughly constant with increasing cluster size for both Spinnaker and Cassandra. This is not surprising since a write only affects the 3 nodes where the value is replicated, regardless of the number of nodes in the cluster. Although the results for read latency are not shown, they also showed similar behavior. Note that we were unable to turn off the write cache of the local disks on EC2. Consequently, the results reported for EC2 are not comparable to the ones reported for our local cluster.

Recall that Zookeeper is not in the critical path of reads and writes in Spinnaker. As a result, we expect a single Zookeeper service to support a Spinnaker cluster with thousands of nodes. However, because it acts as a centralized coordinator, Zookeeper does impose a limit Spinnaker's scaling. In contrast, Cassandra has no centralized coordinator and is inherently more scalable. Eventual consistency also enables Cassandra to scale across datacenters, which is something Spinnaker was not designed to do.

## D.3 Mixed Reads and Writes

In this experiment, we compared the performance of Spinnaker and Cassandra on a mixed workload of reads and writes using 4KB values. For reads, Spinnaker's consistent and timeline reads were compared to Cassandra's quorum and weak reads. For writes, Cassandra's quorum writes were used so its durability was the same as Spinnaker.

Figure 12 shows the average latency of the mixed workload as the percentage of writes was increased. The load was fixed at 2 client threads. As expected, latency increased as the percentage of writes increased. With 10% writes, the latency of the mix with Spinnaker's consistent read was about 10% better than the mix with Cassandra's quorum read. But with 50% writes, Cassandra was roughly 7% better. With weaker consistency, the latency of the mix with Spinnaker's timeline read was 2% to 10% worse than the mix with Cassandra's weak reads across the board. These results reaffirm that, compared to an eventually consistent datastore, Spinnaker can support a stronger consistency model with only a small loss in performance.

## D.4 Using an SSD for Logging

A solid-state disk (SSD) is an attractive option for a logging device since an SSD can provide durable writes with very low latencies. Although the per GB cost of an SSD is currently much higher than for a magnetic disk, only a small amount of space is typically needed for logging, since logs are regularly rolled over.

Figure 13 shows the average latency of a write in Spinnaker and Cassandra under the same settings as before except that each of the nodes in the cluster used a FusionIO ioXtreme drive with 80GB of non-volatile NAND flash storage to store its log. As shown, the average latency of a write improved dramatically on both datastores to 6 msec or less in most cases. Compared to the write results presented earlier, the relative improvement was larger for Spinnaker because it is more sensitive to logging performance.

In addition to improving performance, note that using an SSD for logging could also help simplify Spinnaker's architecture. For example, a per-node shared log file is no longer necessary with an SSD to achieve acceptable performance. This is because, unlike a magnetic disk, an SSD can support multiple log files without incurring a seek penalty. With separate log files, the bookkeeping needed to track the key range that each log record belongs to could be eliminated.

## D.5 Conditional Put

In this experiment, we ran the write workload described earlier, inserting a random 4KB value for each key. Then clients used the conditional put API to atomically replace existing values with a new random value. The average latency of Spinnaker's conditional put with increasing load is shown in Figure 14. As expected, conditional put performed marginally worse than a regular put. This is because a conditional put has to read a version number and perform a comparison before writing the new value.

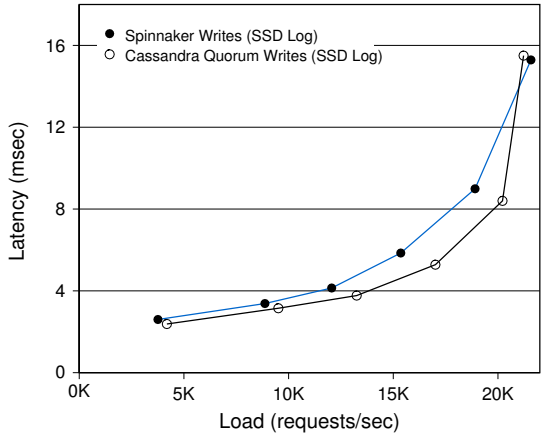

Figure 13: Average write latency using an SSD for logging.

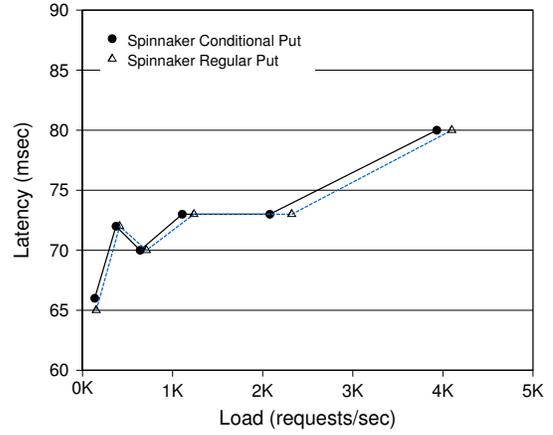

Figure 14: Conditional put vs. regular put in Spinnaker.

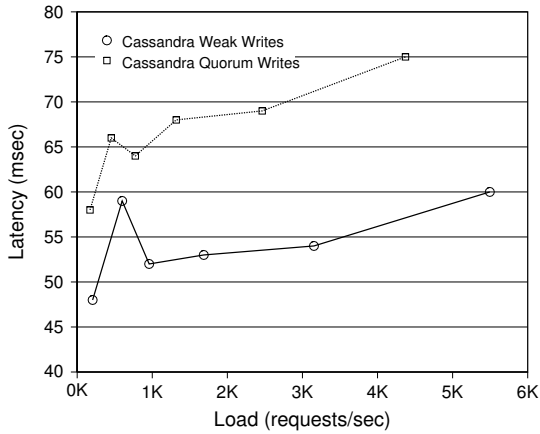

Figure 15: Weak vs. quorum writes in Cassandra.

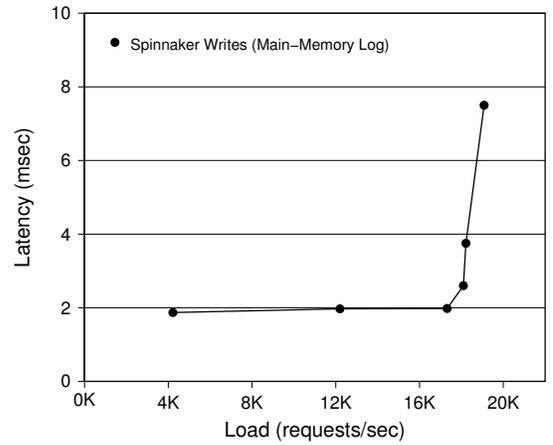

Figure 16: Average write latency with a main memory log in Spinnaker.

## D.6 Lower Durability Guarantees

### D.6.1 Weak Writes in Cassandra

A *weak write* in Cassandra only guarantees that the write has been logged to 1 disk before returning to the client. It should be clear that the durability of a weak write is poorer than a quorum write in Cassandra. With weak writes, a single node or disk failure can cause committed data to be lost. Anecdotally, we have heard that most applications do not use weak writes in Cassandra for this reason. Nonetheless, out of curiosity, we re-ran our write experiments to measure the performance of Cassandra's weak writes. Figure 15 compares the average latency of Cassandra's weak write to its quorum write. As shown, the latency of Cassandra's quorum write was about 40% to 50% slower than its weak write.

### D.6.2 Main Memory Logs in Spinnaker

Note that it is possible to provide strong consistency and weak durability at the same time. To explore this possibility, we experimented with a modified version of Spinnaker that allowed a write to be committed after being written to 2 out of 3 main memory logs. A background thread wrote the main memory log to disk to prevent it from growing too large.

Unlike Cassandra's weak write, a single node failure will not cause committed data to be lost in this approach. However, in a correlated power outage where all the nodes in a cohort go down simultaneously, a small number of committed writes could be lost. For certain applications, this may be an acceptable risk to take in exchange for improved performance. We believe that this durability setting could be suitable for many Web 2.0 applications.

Figure 16 shows the average latency of a write using main memory logs. As shown, committing to 2 out of 3 main memory logs improved Spinnaker's write latency to about 2 msec. We believe this latency could be further reduced with some improvements to Spinnaker's messaging layer.